\documentclass[sigconf]{acmart}

\usepackage{multicol,graphicx,mathrsfs,pifont,amscd,latexsym,color,fancyhdr,url,longtable, pifont, subfigure,epstopdf,setspace,multirow,colortbl,tabularx, booktabs, bbm,listings, balance,color,indentfirst,algorithmic,algorithm,enumitem, subfigure}
\usepackage{amsmath}

\begin{document}

\copyrightyear{2019}
\acmYear{2019} 
\acmConference[WWW '19]{Proceedings of the 2019 World Wide Web Conference}{May 13--17, 2019}{San Francisco, CA, USA}
\acmBooktitle{Proceedings of the 2019 World Wide Web Conference (WWW '19), May 13--17, 2019, San Francisco, CA, USA}
\acmDOI{10.1145/3308558.3313447}
\acmISBN{978-1-4503-6674-8/19/05}

\title{Unbiased LambdaMART: An Unbiased Pairwise Learning-to-Rank Algorithm}

\begin{abstract}

Recently a number of algorithms under the theme of `unbiased learning-to-rank' have been proposed, which can reduce position bias, the major type of bias in click data, and train a high-performance ranker with click data in learning-to-rank. Most of the existing algorithms, based on the inverse propensity weighting (IPW) principle, first estimate the click bias at each position, and then train an unbiased ranker with the estimated biases using a learning-to-rank algorithm. However, there has not been a method for unbiased pairwise learning-to-rank that can simultaneously conduct debiasing of click data and training of a ranker using a pairwise loss function. In this paper, we propose a novel framework to accomplish the goal and apply this framework to the state-of-the-art pairwise learning-to-rank algorithm, LambdaMART. Our algorithm named Unbiased LambdaMART can jointly estimate the biases at click positions and the biases at unclick positions, and learn an unbiased ranker. Experiments on benchmark data show that Unbiased LambdaMART can significantly outperform existing algorithms by large margins. In addition, an online A/B Testing at a commercial search engine shows that Unbiased LambdaMART can effectively conduct debiasing of click data and enhance relevance ranking.
\end{abstract}



\author{Ziniu Hu}
\authornote{This work was done when the first author was an intern at ByteDance AI Lab.}
\affiliation{
  \institution{University of California, Los Angeles, USA}}
\email{bull@cs.ucla.edu} 
\author{Yang Wang, Qu Peng, Hang Li}
\affiliation{
  \institution{ByteDance AI Lab, Beijing, China}}
\email{{wangyang.127, pengqu, lihang.lh}@bytedance.com}

\maketitle

\section{Introduction}\label{sec:introduction}

Learning-to-rank, which refers to machine learning techniques on automatically constructing a model (ranker) from data for ranking in search, has been widely used in current search systems. Existing algorithms can be categorized into pointwise, pairwise, and listwise approaches according to the loss functions they utilize~\cite{DBLP:series/synthesis/2014Li, DBLP:journals/ieicet/Li11, DBLP:journals/ftir/Liu09}. Among the proposed algorithms, LambdaMART is a state-of-the-art algorithm~\cite{DBLP:journals/ir/WuBSG10, from-ranknet-to-lambdarank-to-lambdamart-an-overview}. The data for training in learning-to-rank is usually labeled by human assessors so far, and the labelling process is often strenuous and costly. This raises the question of whether it is possible to train a ranker by using click data collected from the same search system. Click data is indicative of individual users' relevance judgments and is relatively easy to collect with low cost. On the other hand, it is also noisy and biased~\cite{DBLP:conf/sigir/JoachimsGPHG05, DBLP:conf/www/YuePR10}. Notably, the orders of documents in the search results affect users' judgments. Users tend to more frequently click documents presented at higher positions, which is called position bias. This has been preventing practical use of click data in learning-to-rank.

Recently a new research direction, referred to as unbiased learning-to-rank, is arising and making progress. Unbiased learning-to-rank aims at eliminating bias in click data, particularly position bias, and making use of the debiased data to train a ranker. Wang et al.~\cite{DBLP:conf/sigir/WangBMN16} and Joachims et al.~\cite{DBLP:conf/wsdm/JoachimsSS17} respectively propose employing the inverse propensity weighting (IPW) principle~\cite{rose:rubi:cent:1983} to learn an `unbiased ranker' from click data. It is proved that the objective function in learning using IPW is an unbiased estimate of the risk function defined on a relevance measure (a pointwise loss). The authors also develop methods for estimating position bias by randomization of search results online. Wang et al.~\cite{DBLP:conf/wsdm/WangGBMN18} further develop a method for estimating position bias from click data offline. More recently Ai et al.~\cite{DBLP:conf/sigir/AiBLGC18} propose a method that can jointly estimate position bias and train a ranker from click data, again on the basis of IPW. In the previous work, the IPW principle is limited to the pointwise setting in the sense that position biases are only defined on click positions.

In this paper, we address the problem of jointly estimating position biases and training a ranker from click data for pairwise learning-to-rank, particularly using a pairwise algorithm, LambdaMART. To do so, we extend the inverse propensity weighting principle to the pairwise setting, and develop a new method for jointly conducting position bias estimation and ranker training.

Specifically, we give a formulation of unbiased learning-to-rank for the pairwise setting and extend the IPW principle. We define position biases as the ratio of the click probability to the relevance probability at each position, as well as the ratio of the unclick probability to the irrelevance probability. This definition takes both the position biases at click positions and those at unclick positions into consideration. We prove that under the extended IPW principle, the objective function becomes an unbiased estimate of risk function defined on pairwise loss functions. In this way, we can learn an unbiased ranker using a pairwise ranking algorithm.

We then develop a method for jointly estimating position biases for both click and unclick positions and training a ranker for pairwise learning-to-rank, called Pairwise Debiasing. The position bias and the ranker can be iteratively learned through minimization of the same objective function. As an instance, we further develop Unbiased LambdaMART\footnote[1]{Code is available at~\url{https://github.com/acbull/Unbiased_LambdaMart}}, an algorithm of learning an unbiased ranker using LambdaMART.

Experiments on the Yahoo learning-to-rank challenge benchmark dataset demonstrate that Unbiased LambdaMART can effectively conduct debiasing of click data and significantly outperform the baseline algorithms in terms of all measures, for example, 3-4\% improvements in terms of NDCG@1. An online A/B Testing at a commercial news search engine, Jinri Toutiao, also demonstrates that Unbiased LambdaMART can enhance the performance of relevance ranking at the search engine.

The contribution of this paper includes the following proposals.
\begin{itemize}
\item A general framework on unbiased learning-to-rank in the pairwise setting, particularly, an extended IPW.
\item Pairwise Debiasing, a method for jointly estimating position bias and training a pairwise ranker.
\item Unbiased LambdaMART, an algorithm of unbiased pairwise learning-to-rank using LambdaMART.
\end{itemize}

\section{Related Work}\label{sec:related}

In this section, we introduce related work on learning-to-rank, click model, and unbiased learning to rank. 

\subsection{Learning-to-Rank}

Learning-to-rank is to automatically construct a ranking model from data, referred to as a ranker, for ranking in search. A ranker is usually defined as a function of feature vector based on a query document pair. In search, given a query, the retrieved documents are ranked based on the scores of the documents given by the ranker. The advantage of employing learning-to-rank is that one can build a ranker without the need of manually creating it, which is usually tedious and hard. Learning-to-rank is now becoming a standard technique for search.  

There are many algorithms proposed for learning-to-rank. The algorithms can be categorized as pointwise approach, pairwise approach, and listwise approach, based on the loss functions in learning~\cite{DBLP:series/synthesis/2014Li, DBLP:journals/ieicet/Li11, DBLP:journals/ftir/Liu09}. The pairwise and listwise algorithms usually work better than the pointwise algorithms~\cite{DBLP:series/synthesis/2014Li}, because the key issue of ranking in search is to determine the orders of documents but not to judge the relevance of documents, which is exactly the goal of the pairwise and listwise algorithms. For example, the pairwise algorithms of RankSVM~\cite{DBLP:conf/kdd/Joachims02, Cao:2006:ARS:1148170.1148205} and LambdaMART~\cite{DBLP:journals/ir/WuBSG10, from-ranknet-to-lambdarank-to-lambdamart-an-overview} are state-of-the-art algorithms for learning-to-rank. 

Traditionally, data for learning a ranker is manually labeled by humans, which can be costly. To deal with the problem, one may consider using click data as labeled data to train a ranker. Click data records the documents clicked by the users after they submit queries, and it naturally represents users' implicit relevance judgments on the search results. The utilization of click data has both pros and cons. On one hand, it is easy to collect a large amount of click data with low cost. On the other hand, click data is very noisy and has position bias, presentation bias, etc. Position bias means that users tend to more frequently click documents ranked at higher positions~\cite{DBLP:conf/sigir/JoachimsGPHG05, DBLP:conf/www/YuePR10}. How to effectively cope with position bias and leverage click data for learning-to-rank thus becomes an important research issue.

\subsection{Click Model}

One direction of research on click data aims to design a click model to simulate users' click behavior, and then estimate the parameters of the click model from data.  It then makes use of the learned click model for different tasks, for example, use them as features of a ranker. 

Several probabilistic models have been developed.  For example, Richardson et al.~\cite{DBLP:conf/www/RichardsonDR07} propose the Position Based Model (PBM), which assumes that a click only depends on the position and relevance of the document.  Craswell et al.~\cite{DBLP:conf/wsdm/CraswellZTR08} develop the Cascade Model (CM), which formalizes the user's behavior in browsing of search results as a sequence of actions. Dupret et al.~\cite{DBLP:conf/sigir/DupretP08} propose the User Browsing Model (UBM), asserting that a click depends not only on the position of a document, but also on the positions of the previously clicked documents. Chapelle et al.~\cite{DBLP:conf/www/ChapelleZ09} develop the Dynamic Bayesian Network Model (DBN), based on the assumption that the user's behavior after a click does not depend on the perceived relevance of the document but on the actual relevance of the document. Borisov et al.~\cite{DBLP:conf/www/BorisovMRS16} develop the Neural Click Model, which utilizes neural networks and vector representations to predict user's click behavior. 
Recently, Kveton et al.~\cite{DBLP:conf/icml/KvetonSWA15} propose a multi-armed bandit learning algorithm of the Cascade Model to identify the $k$ most attractive items in the ranking list. Li et al.~\cite{DBLP:conf/kdd/LiAKMVW18} makes use of click models to evaluate the performance of ranking model offline.

Click models can be employed to estimate position bias and other biases, as well as query document relevance. They are not designed only for the purpose of debiasing and thus could be sub-optimal for the task. In our experiments, we use the click models to generate synthetic click datasets for evaluating our proposed unbiased learning-to-rank algorithm offline.

\subsection{Unbiased Learning-to-Rank}

Recently, a new direction in learning-to-rank, referred to as unbiased learning-to-rank, is arising and making progress. 
The goal of unbiased learning-to-rank is to develop new techniques to conduct debiasing of click data and leverage the debiased click data in training of a ranker\cite{DBLP:conf/cikm/AiMLC18}.

Wang et al.~\cite{DBLP:conf/sigir/WangBMN16} apply unbiased learning-to-rank to personal search. They conduct randomization of search results to estimate query-level position bias and adjust click data for training of a ranker in personal search on the basis of inverse propensity weighting (IPW)~\cite{rose:rubi:cent:1983}. Joachims et al.~\cite{DBLP:conf/wsdm/JoachimsSS17} theoretically prove that with the inverse propensity weighting (IPW) principle, one can obtain an unbiased estimate of a risk function on relevance in learning-to-rank.  They also utilize online randomization to estimate position bias and perform training of a RankSVM model. Wang et al.~\cite{DBLP:conf/wsdm/WangGBMN18} employ a regression-based EM algorithm to infer position bias by maximizing the likelihood of click data. The estimated position bias is then utilized in learning of LambdaMART. Recently, Ai et al.~\cite{DBLP:conf/sigir/AiBLGC18} design a dual learning algorithm which can jointly learn an unbiased propensity model for representing position bias and an unbiased ranker for relevance ranking, by optimizing two objective functions. Both models are implemented as neural networks. Their method is also based on IPW, while the loss function is a pointwise loss function.

Our work mainly differs from the previous work in the following points:\begin{itemize}
    \item In previous work, position bias (propensity) is defined as the observation probability, and thus IPW is limited to the pointwise setting in which the loss function is pointwise and debiasing is performed at a click position each time. In this work, we give a more general definition of position bias (propensity), and extend IPW to the pairwise setting, in which the loss function is pairwise and debiasing is carried out at both click positions and unclick positions.
    
    \item In previous work, estimation of position bias either relies on randomization of search results online, which can hurt user experiences~\cite{DBLP:conf/sigir/WangBMN16, DBLP:conf/wsdm/JoachimsSS17}, or resorts to separate learning of a propensity model from click data offline, which can be suboptimal to relevance ranking~\cite{DBLP:conf/wsdm/WangGBMN18,DBLP:conf/sigir/AiBLGC18}. In this paper, we propose to jointly conduct estimation of position bias and learning of a ranker through minimizing one objective function. We further apply this framework to the state-of-the-art LambdaMART algorithm.
\end{itemize}

\section{Framework}\label{sec:ul2r}

In this section, we give a general formulation of unbiased learning-to-rank, for both the pointwise and pairwise settings. We also extend the inverse propensity weighting principle to the pairwise setting.
\begin{table}[t]
\caption{A summary of notations.}
\label{tab:notations} 
\begin{center}
\begin{tabular}{|l|p{4.9cm}|} \hline
$q$, $D_q$ & query $q$ and documents $D_q$ of $q$ \\ \hline
$i$, $d_i$, $x_i$, $r_i$, $c_i$ & $i$-th (representing the position by original ranker where click data is collected) document $d_i$ in $D_q$ with feature vector $x_i$, relevance information $r_i$ (1/0) and click information $c_i$ (1/0) \\ \hline
$I_q = \{(d_i,d_j)\}$ & set of pairs of documents of $q$, in which $d_i$ is more relevant or more clicked than $d_j$ \\ \hline
$C_q$, $\mathcal{D} = \{ (q,D_q,C_q) \}$ & click information $C_q$ of $D_q$ and click data set $\mathcal{D}$ for all queries \\
\hline
\end{tabular}
\end{center}
\end{table}
\subsection{Pointwise Unbiased Learning-to-Rank}

In learning-to-rank, given a query document pair denoted as $x$, the ranker $f$ assigns a score to the document. The documents with respect to the query are then ranked in descending order of their scores. Traditionally, the ranker is learned with labeled data. In the pointwise setting, the loss function in learning is defined on a single data point $x$.

Let $q$ denote the query and $D_q$ the set of documents associated with $q$. Let $d_i$ denote the $i$-th document in $D_q$ and $x_i$ the feature vector of $q$ and $d_i$  (see Table 1). Let $r^{+}_i$ and $r^{-}_i$ represent that $d_i$ is relevant and irrelevant respectively (i.e., $r_i = 1$ and $r_i = 0$). For simplicity we only consider binary relevance here and one can easily extend it to the multi-level relevance case. The risk function in learning is defined as
\begin{align}
    R_{_{rel}}(f) =& \int L(f(x_i), r^+_i) \ dP(x_i, r^+_i)
\end{align}
where $f$ denotes a ranker, $L(f(x_i), r^+_{i})$ denotes a pointwise loss function based on an IR measure~\cite{DBLP:conf/wsdm/JoachimsSS17} and $P(x_i, r^+_i)$ denotes the probability distribution on $x_i$ and $r^+_i$. Most ranking measures in IR only utilize relevant documents in their definitions, and thus the loss function here is defined on relevant documents with label $r^+_i$. Furthermore, the position information of documents is omitted from the loss function for notation simplicity. 

Suppose that there is a labeled dataset in which the relevance of documents with respect to queries is given. One can learn a ranker $\hat{f}_{_{rel}}$ through 
the minimization of the empirical risk function (objective function) as follows.
\begin{align}
    \hat{f}_{_{rel}} =&\mathop{\arg\min_f} \sum_{\substack{q}}\sum_{\substack{d_i \in D_q}} \ L(f(x_i), r^+_{i})
\end{align}

One can also consider using click data as relevance feedbacks from users, more specifically, viewing clicked documents as relevant documents and unclicked documents as irrelevant documents, and training a ranker with a click dataset. This is what we call `biased learning-to-rank', because click data has position bias, presentation bias, etc. Suppose that there is a click dataset in which the clicks of documents with respect to queries by an original ranker are recorded. For convenience, let us assume that document $d_i$ in $D_q$ is exactly the document ranked at position $i$ by the original ranker. Let $c^{+}_i$ and $c^{-}_i$ represent that document $d_i$ is clicked and unclicked in the click dataset respectively (i.e., $c_i = 1$ and $c_i = 0$). The risk function and minimization of empirical risk function can be defined as follows.
\begin{align}
    R_{_{click}}(f) =& \int L(f(x_i), c^+_i) \ dP(x_i, c^+_i)\\
    \hat{f}_{_{click}} =&\mathop{\arg\min_f} \sum_{\substack{q}} \sum_{\substack{d_i \in D_q}} \ L(f(x_i), c^+_{i})
\end{align}
The loss function is defined on clicked documents with label $c^+_i$. The ranker $\hat{f}_{_{click}}$ learned in this way is biased, however. 

Unbiased learning-to-rank aims to eliminate the biases, for example position bias, in the click data and train a ranker with the debiased data. The training of ranker and debiasing of click data can be performed simultaneously or separately. The key question is how to fill the gap between click and relevance, that is, $P(c^{+}_i|x_i)$ and $P(r^{+}_i|x_i)$.  Here we assume that the click probability is proportional to the relevance probability at each position, where the ratio $t^{+}_i >0$ is referred to as bias at a click position $i$.
\begin{align}
  P(c^{+}_i|x_i)   &=  t^{+}_i P(r^{+}_i|x_i) \label{con:click}
\end{align}
There are $k$ ratios corresponding to $k$ positions. The ratios can be affected by different types of bias, but in this paper, we only consider position bias.

We can conduct learning of an unbiased ranker $\hat{f}_{_{unbiased}}$, through minimization of the empirical risk function as follows.
\begin{align}
    R_{_{unbiased}}(f) &= \int \frac{L(f(x_i), c^+_i)}{t^{+}_i} \ dP(x_i, c^+_i)\\
    &= \int \frac{L(f(x_i), c^+_i)}{\frac{P(c^{+}_i|x_i)}{P(r^{+}_i|x_i)}} \ dP(x_i, c^+_i)\\
    &= \int L(f(x_i), c^+_i) \ dP(x_i, r^+_i) \\
    &= \int L(f(x_i), r^+_i) \ dP(x_i, r^+_i) = R_{_{rel}}(f)\label{con:change}\\
    \hat{f}_{_{unbiased}} &= \mathop{\arg\min_f} \sum_{\substack{q}}\sum_{\substack{d_i \in D_q}} \ \frac{L(f(x_i), c^+_i)}{t^{+}_i}  \label{con:ipw}
\end{align}
In (\ref{con:change}) click label $c^+_i$ in the loss function is replaced with relevance label $r^+_i$, because after debiasing click implies relevance.

One justification of this method is that $R_{unbiased}$ is in fact an unbiased estimate of $R_{rel}$. This is the so-called inverse propensity weighting (IPW) principle proposed in previous work. That is to say, if we can properly estimate position bias (ratio) $t^{+}_i$, then we can reliably train an unbiased ranker $\hat{f}_{_{unbiased}}$.

An intuitive explanation of position bias (ratio) $t^{+}_i$ can be found in the following relation, under the assumption that a clicked document must be relevant ($c^{+} \Rightarrow r^{+}$). 
\begin{align}
    t^{+}_i = \frac{P(c^{+}_i|x_i)}{P(r^{+}_i|x_i)} = \frac{P(c^{+}_i, r^{+}_i|x_i)}{P(r^{+}_i|x_i)} = P(c^{+}_i| r^{+}_i,x_i)  \label{con:positive}
\end{align}
It means that $t^{+}_i$ represents the conditional probability of how likely a relevant document is clicked at position $i$ after examination of the document. In the original IPW, $t^{+}_i$ is defined as the observation probability that the user examines the document at position $i$ before clicking the document~\cite{DBLP:conf/wsdm/JoachimsSS17,DBLP:conf/wsdm/WangGBMN18}, which is based on the same assumption as (\ref{con:positive}). 

\subsection{Pairwise Unbiased Learning-to-Rank}

In the pairwise setting, the ranker $f$ is still defined on a query document pair $x$, and the loss function is defined on two data points $x_i$ and $x_j$. Traditionally, the ranker is learned with labeled data.

Let $q$ denote a query. Let $d_i$ and $d_j$ denote the $i$-th and $j$-th documents with respect to query $q$. Let $x_i$ and $x_j$ denote the feature vectors from $d_i$ and $d_j$ as well as $q$. Let $r^{+}_i$ and $r^{-}_j$ represent that document $d_i$ and document $d_j$ are relevant and irrelevant respectively. Let $I_q$ denote the set of document pairs $(d_i,d_j)$ where $d_i$ is relevant and $d_j$ is irrelevant. For simplicity we only consider binary relevance here and one can easily extend it to the multi-level relevance case. The risk function and the minimization of empirical risk function are defined as
\begin{align}
  R_{_{rel}}(f) = & \int L(f(x_i), r^+_i, f(x_j), r^-_j) \ dP(x_i, r^+_i, x_j, r^-_j) \\
  \hat{f}_{_{rel}} = &\mathop{\arg\min_f} \sum_{\substack{q}}\sum_{\substack{(d_i,d_j)\in I_q}} \ L(f(x_i), r^+_i, f(x_j), r^-_j)
\end{align}
where $L(f(x_i), r^+_i, f(x_j), r^-_j)$ denotes a pairwise loss function. 

One can consider using click data to directly train a ranker, that is, to conduct `biased learning-to-rank'. Let $c^{+}_i$ and $c^{-}_j$ represent that document $d_i$ and document $d_j$ are clicked and unclicked respectively. Let $I_q$ denote the set of document pairs $(d_i,d_j)$ where $d_i$ is clicked and $d_j$ is unclicked. The risk function and minimization of empirical risk function can be defined as follows.
\begin{align}
    R_{_{click}}(f) = &\int L(f(x_i), c^+_i, f(x_j), c^-_j)  \ dP(x_i, c^+_i, x_j, c^-_j)\\
    \hat{f}_{_{click}} = &\mathop{\arg\min_f} \sum_{\substack{q}}\sum_{\substack{(d_i,d_j)\in I_q}} \ L(f(x_i), c^+_i, f(x_j), c^-_j)
\end{align}
The ranker $\hat{f}_{_{click}}$ is however biased.

Similar to the pointwise setting, we consider dealing with position bias in the pairwise setting and assume that the click probability is proportional to the relevance probability at each position and the unclick probability is proportional to the irrelevance probability at each position. The ratios $t^{+}_i >0$ and $t^{-}_j >0$ are referred to as position biases at a click position $i$ and an unclick position $j$.
\begin{align}
    P(c^{+}_i|x_i)  &=  t^{+}_i P(r^{+}_i|x_i) \label{con:click1}\\
    P(c^{-}_j|x_j)  &=  t^{-}_j P(r^{-}_i|x_j) \label{con:click2}
\end{align}
There are $2k$ position biases (ratios) corresponding to $k$ positions.

We can conduct learning of an unbiased ranker $\hat{f}_{_{unbiased}}$, through minimization of the empirical risk function as follows.
\begin{normalsize}
\begin{align}
    R_{_{unbiased}}(f)  = &\int \frac{L(f(x_i), c^+_i, f(x_j), c^-_j)}{t^{+}_i \cdot t^{-}_j} \ dP(x_i, c^+_i, x_j, c^-_j) \label{con:depart}\\
     = & \int\int\frac{L(f(x_i), c^+_i, f(x_j), c^-_j)dP(c^{+}_i,x_i)dP(c^{-}_j,x_j)}{\frac{P(c^{+}_i|x_i)P(c^{-}_j|x_j)}{P(r^{+}_i|x_i)P(r^{-}_i|x_j)}} \\
     = & \int\int L(f(x_i), c^+_i, f(x_j), c^-_j)dP(r^{+}_i,x_i)dP(r^{-}_i,x_j)\\
     = & \int L(f(x_i), r^+_i, f(x_j), r^-_j)dP(x_i, r^+_i, x_j, r^-_j) \label{con:output}\\
     = & R_{_{rel}}(f)\\
    \hat{f}_{_{unbiased}} =& \mathop{\arg\min_f} \sum_{\substack{q}}\sum_{\substack{(d_i,d_j) \in I_q}} \ \frac{L(f(x_i), c^+_i, f(x_j),c^-_j)}{t^{+}_i \cdot t^{-}_j}  \label{con:way} 
\end{align}
\end{normalsize}
In (\ref{con:depart}) it is assumed that relevance and click at position $i$ are independent from those at position $j$. (Experimental results show that the proposed Unbiased LambdaMART works very well under this assumption, even one may think that it is strong.) In (\ref{con:output}), click labels $c^+_i$ and $c^-_j$ are replaced with relevance labels $r^+_i$ and $r^-_j$ because after debiasing click implies relevance and unclick implies irrelevance. 

One justification of this method is that $R_{unbiased}$ is an unbiased estimate of $R_{rel}$. Therefore, if we can accurately estimate the position biases (ratios), then we can reliably train an unbiased ranker $\hat{f}_{_{unbiased}}$. This is an extension of the inverse propensity weighting (IPW) principle to the pairwise setting.

Position bias (ratio) $t^{+}_i$ has the same explanation as that in the pointwise setting. An explanation of position bias (ratio) $t^{-}_j$ is that it represents the reciprocal of the conditional probability of how likely an unclicked document is irrelevant at position $j$, as shown below. 
\begin{align}
    t^{-}_j = \frac{P(c^{-}_j|x_j)}{P(r^{-}_i|x_j)} = \frac{P(c^{-}_j|x_j)}{P(r^{-}_j,c^{-}_j|x_j)} = \frac{1}{P(r^{-}_j|c^{-}_j, x_j)} \label{con:negative}
\end{align}
It is under the assumption that an irrelevant document must be unclicked ($r^{-} \Rightarrow c^{-}$), which is equivalent to ($c^{+} \Rightarrow r^{+}$). Note that $t^{-}_j$ is not a probability and it has a different interpretation from $t^{+}_i$. The unclicked document $j$ can be either examined or unexamined. Thus, in the extended IPW the condition on examination of document in the original IPW is dropped.

\section{Approach}\label{sec:approach}

In this section, we present Pairwise Debiasing as a method of jointly estimating position bias and training a ranker for unbiased pairwise learning-to-rank. Furthermore, we apply Pairwise Debiasing on LambdaMart and describe the learning algorithm of Unbiased LambdaMART.


\subsection{Learning Strategy}

We first give a general strategy for pairwise unbiased learning-to-rank, named Pairwise Debiasing. 

A key issue of unbiased learning-to-rank is to accurately estimate position bias. Previous work either relies on randomization of search results online, which can hurt user experiences~\cite{DBLP:conf/sigir/WangBMN16, DBLP:conf/wsdm/JoachimsSS17}, or resorts to a separate learning of position bias from click data offline, which can be suboptimal to the ranker~\cite{DBLP:conf/wsdm/WangGBMN18,DBLP:conf/sigir/AiBLGC18}. In this paper, we propose to simultaneously conduct estimation of position bias and learning of a ranker offline through minimizing the following regularized loss function (objective function).

\begin{small}
\begin{align}
    \min_{f, t^{+}, t^{-}} & \ \mathcal{L}(f, t^{+}, t^{-})\\ 
 =  \min_{f, t^{+}, t^{-}} & \ \sum_{q}\sum_{\substack{ (d_i,d_j) \in I_q}} \ \frac{L(f(x_i), c^+_i, f(x_j), c^-_j)}{t^{+}_i \cdot t^{-}_j} + ||t^{+}||^p_p + ||t^{-}||^p_p \label{opti} \\
                         & s.t. \ t^{+}_1 = 1 , \ t^{-}_1 = 1
\end{align}
\end{small}
where $f$ denotes a ranker, $t^{+}$ and $t^{-}$ denote position biases (ratios) at all positions, $L$ denotes a pairwise loss function, $||\cdot||^p_p$ denotes $L_p$ regularization. Because the position biases are relative values with respect to positions, to simplify the optimization process we fix the position biases of the first position to $1$ and only learn the (relative) position biases of the rest of the positions. Here $p \in [0,+\infty)$ is a hyper-parameter. The higher the value of $p$ is, the more regularization we impose on the position biases.

In the objective function, the position biases $t^{+}$ and $t^{-}$ are inversely proportional to the pairwise loss function $L(f(x_i), c^+_i, f(x_j), c^-_j)$, and thus the estimated position biases will be high if the losses on those pairs of positions are high in the minimization. The position biases are regularized and constrained to avoid a trivial solution of infinity.

It would be difficult to directly optimize the objective function in (\ref{opti}). We adopt a greedy approach to perform the task. Specifically, for the three optimization variables $f$, $t^{+}$, $t^{-}$, we iteratively optimize the objective function $\mathcal{L}$ with respect to one of them with the others fixed; we repeat the process until convergence. 

\subsection{Estimation of position bias ratios}

Given a fixed ranker, we can estimate the position biases at all positions. There are in fact closed form solutions for the estimation.

The partial derivative of objective function $\mathcal{L}$ with respect to position bias $t^{+}$ is 
\begin{small}
\begin{align}
    \frac{\partial \mathcal{L}(f^*, t^{+}, (t^{-})^*)}{\partial t_i^{+}} = \sum_{q}\sum_{\substack{j:(d_i,d_j) \in I_q}} \frac{L(f^*(x_i), c^+_i, f^*(x_j), c^-_j)}{-(t_i^{+})^2 \cdot (t^{-}_j)^*} + p \cdot (t_i^{+})^{p-1}
\end{align}
\end{small}
Thus, we have\footnote{The derivation is based on the fact $p \in (0,+\infty)$. The result is then extended to the case of $p=0$.}
\begin{align}
    \mathop{\arg\min_{t^{+}_i}} \mathcal{L}(f^*, t^{+}, (t^{-})^*) =  \left[\sum_{q}\sum_{\substack{j:(d_i,d_j) \in I_q}}\frac{L(f^*(x_i), c^+_i, f^*(x_j), c^-_j)}{p \cdot (t^{-}_j)^*}\right]^{\frac{1}{p+1}}
\end{align}
\begin{align}   
    t^{+}_i
              &= \left[\frac{\sum_{q}\sum_{j:\substack{(d_i,d_j) \in I_q}} (L(f^*(x_i), c^+_i, f^*(x_j), c^-_j) \ / \ (t^{-}_j)^*)}{\sum_{q}\sum_{k:\substack{(d_1,d_k) \in I_q}} (L(f^*(x_1), c^+_1, f^*(x_k), c^-_k) \ / \ (t^{-}_k)^*)}\right]^{\frac{1}{p+1}} \label{con:update}
\end{align}
In (\ref{con:update}) the result is normalized to make the position bias at the first position to be $1$.

Similarly, we have
\begin{align}
    t^{-}_j
              &= \left[\frac{\sum_{q}\sum_{\substack{i:(d_i,d_j) \in I_q}} (L(f^*(x_i), c^+_i, f^*(x_j), c^-_j) \ / \ (t^{+}_i)^*)}{\sum_{q}\sum_{\substack{k:(d_k,d_1) \in I_q}} (L(f^*(x_k), c^+_k, f^*(x_1), c^-_1) \ / \ (t^{+}_k)^*)}\right]^{\frac{1}{p+1}} \label{con:update2}
\end{align}
In this way, we can estimate the position biases (ratios) $t^+$ and $t^-$ in one step given a fixed ranker $f^*$. Note that the method here, referred to as Pairwise Debiasing, can be applied to any pairwise loss function. In this paper, we choose to apply the pairwise learning-to-rank algorithm LambdaMART.

\subsection{Learning of Ranker}

Given fixed position biases, we can learn an unbiased ranker. The partial derivative of $\mathcal{L}$ with respect to $f$ can be written in the following general form.
\begin{small}
\begin{align}
    \frac{\partial \mathcal{L}(f, (t^{+})^*, (t^{-})^*)}{\partial f} = \sum_{q}\sum_{\substack{(d_i,d_j) \in I_q}} \frac{1}{(t^{+}_i)^*\cdot (t^{-}_j)^*} \frac{\partial L(f(x_i), c^+_i, f(x_j), c^-_j)}{\partial f}
\end{align}
\end{small}

We employ LambdaMART to train a ranker. LambdaMART~\cite{DBLP:conf/nips/BurgesRL06, DBLP:journals/ir/WuBSG10} employs gradient boosting or MART~\cite{Friedman00greedyfunction} and the gradient function of the loss function called lambda function. Given training data, it performs minimization of the objective function using the lambda function.

In LambdaMART, the lambda gradient $\lambda_i$ of document $d_i$ is calculated using all pairs of the other documents with respect to the query.

\begin{align}
    \lambda_{i} = & \sum_{j:(d_i,d_j) \in I_q}{\lambda_{ij} -\sum_{j:(d_j,d_i) \in I_q}\lambda_{ji}} \label{con:lambda}
\end{align}

\begin{align} 
    \lambda_{ij}
                =&\frac{-\sigma}{1+e^{\sigma(f(x_i)-f(x_j))}}\left|  \Delta Z_{ij} \right| 
                \label{con:lambda2}
\end{align}
where $\lambda_{ij}$ is the lambda gradient defined on a pair of documents $d_i$ and $d_j$, $\sigma$ is a constant with a default value of 2, $f(x_i)$ and $f(x_j)$ are the scores of the two documents given by LambdaMART, $\Delta Z_{ij}$ denotes the difference between NDCG\cite{DBLP:journals/tois/JarvelinK02} scores if documents $d_i$ and $d_j$ are swapped in the ranking list. 

Following the discussion above, we can make an adjustment on the lambda gradient $\tilde{\lambda}_{i}$ with the estimated position biases:
\begin{align}
\tilde{\lambda}_{i} = & \sum_{j:(d_i,d_j) \in I_q}{\tilde{\lambda}_{ij} -\sum_{j:(d_j,d_i) \in I_q}} \label{con:lambda3} 
\tilde{\lambda}_{ji} \\ 
    \tilde{\lambda}_{ij} = & \frac{\lambda_{ij}}{(t^{+}_i)^* \cdot (t^{-}_j)^*} \label{con:lambda4}
\end{align}

Thus, by simply replacing the lambda gradient $\lambda_i$ in LambdaMART with the adjusted lambda gradient $\tilde{\lambda}_{i}$, we can reliably learn an unbiased ranker with the LambdaMART algorithm. We call the algorithm Unbiased LambdaMART.

Estimation of position biases in (\ref{con:update}) and (\ref{con:update2}) needs calculation of the loss function $L_{ij}=L(f(x_i), c^+_i, f(x_j), c^-_j)$. For LambdaMART the loss function can be derived from (\ref{con:lambda2}) as follows.
\begin{align}  
    L_{ij} = \log(1+e^{-\sigma{(f(x_i)-f(x_j))}}) \ \left|  \Delta Z_{ij} \right|  \label{con:loss}
\end{align}

\subsection{Learning Algorithm}

The learning algorithm of Unbiased LambdaMART is given in Algorithm \ref{alg:Framwork}. The input is a click dataset $\mathcal{D}$. The hyper-parameters are regularization parameter $p$ and total number of boosting iterations $M$. The output is an unbiased ranker $f$ and estimated position biases at all positions $t^+$, $t^-$. As is outlined in Algorithm~\ref{alg:Framwork}, Unbiased LambdaMART iteratively calculates adjusted lambda gradient in line \ref{alg:calculate}, re-trains a ranker with the gradients in line \ref{alg:train}, and re-estimates position biases in line \ref{alg:estimate}.  The time complexity of Unbiased LambdaMART is the same as that of LambdaMART.

\begin{algorithm}[htb] 
\caption{Unbiased LambdaMART} 
\label{alg:Framwork} 
\begin{algorithmic}[1] 
\REQUIRE
click dataset $\mathcal{D}=\{(q,D_q,C_q)\}$; 
hyper-parameters $p$, $M$;
\ENSURE
unbiased ranker $f$; 
position biases (ratios) $t^{+}$ and $t^{-}$; 
\STATE Initialize all position biases (ratios) as 1; 
\FOR {$m=1$ to $M$}
    \FOR {each query $q$ and each document $d_i$ in $D_q$}
	    \STATE  Calculate $\tilde{\lambda}_i$ with $(t^{+})^*$ and $(t^{-})^*$ using (\ref{con:lambda3}) and (\ref{con:lambda4}); \label{alg:calculate}
	\ENDFOR
     \STATE  Re-train ranker $f$ with $\tilde{\lambda}$ using LambdaMART algorithm \label{alg:train}
	\STATE  Re-estimate position biases (ratios) $t^{+}$ and $t^{-}$ with ranker $f^*$ using (\ref{con:update}) and (\ref{con:update2}) \label{alg:estimate} \
\ENDFOR
\RETURN $f$, $t^{+}$, and $t^{-}$; 
\end{algorithmic} 
\end{algorithm}

\section{Experiments}\label{sec:evaluation}
In this section, we present the results of two experiments on our proposed algorithm Unbiased LambdaMART. One is an offline experiment on a benchmark dataset, together with empirical analysis on the effectiveness, generalizability, and robustness of the algorithm. The other experiment is an online A/B testing at a commercial search engine.


\subsection{Experiment on Benchmark Data}

We made use of the Yahoo!~learning-to-rank challenge dataset\footnote{\url{http://webscope.sandbox.yahoo.com}} to conduct an experiment. The Yahoo dataset is one of the largest benckmark dataset for learning-to-rank. It consists of 29921 queries and 710k documents. Each query document pair is represented by a 700-dimensional feature vector manually assigned with a label denoting relevance at 5 levels~\cite{DBLP:journals/jmlr/ChapelleC11}.

There is no click data associated with the Yahoo dataset. We followed the procedure in ~\cite{DBLP:conf/sigir/AiBLGC18} to generate synthetically click data from the Yahoo dataset for offline evaluation.\footnote{We plan to release the synthetically generated click data as well as the source code of Unbiased LambdaMART.}

We chose NDCG at position 1,3,5,10 and MAP as evaluation measures in relevance ranking.

\subsubsection{Click Data Generation}


The click data generation process in ~\cite{DBLP:conf/sigir/AiBLGC18} is as follows. First, one trains a Ranking SVM model using 1\% of the training data with relevance labels, and uses the model to create an initial ranking list for each query. Next, one samples clicks from the ranking lists by simulating the browsing process of users. The position-based click model (PBM) is utilized. It assumes that a user decides to click a document according to probability $P(c^+_i) = P(o_i^+)P(r_i^+)$.  Here $P(o_i^+)$ and $P(r_i^+)$ are the observation probability and relevance probability respectively.

The probability of observation $P(o_i^+)$ is calculated by
$$P(o_i^+|x_i) = \rho_{i}^{\theta}$$
where $\rho_{i}$ represents position bias at position $i$ and $\theta \in [0, +\infty]$ is a parameter controlling the degree of position bias. 
The position bias $\rho_{i}$ is obtained from an eye-tracking experiment in ~\cite{DBLP:conf/sigir/JoachimsGPHG05} and the parameter $\theta$ is set as 1 by default. 

The probability of relevance $P(r_i^+)$ is calculated by $$P(r_i^+) = \epsilon + (1-\epsilon)\frac{2^y-1}{2^{y_{max}}-1}$$ where $y \in [0, 4]$ represents a relevance level and $y_{max}$ is the highest level of 4. The parameter $\epsilon$ represents click noise due to that irrelevant documents ($y=0$) are incorrectly perceived as relevant documents ($y>0$) , which is set as 0.1 by default. 

\begin{table*}[ht]
\caption{Comparison of different unbiased learning-to-rank methods.} 
\label{tab:simulate} 
\Large
\begin{tabular}{p{3cm}<{\centering}|p{4.6cm}<{\centering}||p{1.2cm}<{\centering}|p{1.6cm}<{\centering}|p{1.6cm}<{\centering}|p{1.6cm}<{\centering}|p{1.6cm}<{\centering}}
\hline
\multicolumn{1}{c|}{Ranker}  & Debiasing Method & MAP & NDCG@1 & NDCG@3 & NDCG@5 & NDCG@10 \\ \hline\hline
                           
\multirow{4}{*}{\textbf{LambdaMART}}         & Labeled Data (Upper Bound)    &0.854& 0.745  & 0.745  & 0.757  & 0.790   \\ \cline{2-7}    
                                    & \textbf{Pairwise Debiasing}   
                                                     &\textbf{0.836}& \textbf{0.717}  & \textbf{0.716}  & \textbf{0.728}  & \textbf{0.764}   \\ \cline{2-7} 
                                    & Regression-EM~\cite{DBLP:conf/wsdm/WangGBMN18}  
                                                     &0.830& 0.685 &  0.684 & 0.700 &  0.743\\ \cline{2-7} 
                                    & Randomization       &0.827 & 0.669  &  0.678 & 0.690  & 0.728   \\ \cline{2-7} 
                                    & Click Data (Lower Bound)     &0.820& 0.658  & 0.669  & 0.672  & 0.716   \\ \hline\hline
                                        
\multirow{4}{*}{DNN}                & Labeled Data  (Upper Bound)   &0.831& 0.677  & 0.685  & 0.705  & 0.737   \\ \cline{2-7}
                                    & Dual Learning Algorithm~\cite{DBLP:conf/sigir/AiBLGC18}            
                                                     &0.828& 0.674  & 0.683  & 0.697  & 0.734  \\ \cline{2-7}
                                    & Regression-EM  & 0.829& 0.676 & 0.684  & 0.699  & 0.736   \\ \cline{2-7} 
                                                                        & Randomization       &0.825& 0.673  & 0.679  & 0.693  & 0.732  \\ \cline{2-7} 
                                    & Click Data (Lower Bound)   &0.819& 0.637  & 0.651  & 0.667  & 0.711   \\ \hline\hline

\multirow{4}{*}{RankSVM}            & Labeled Data (Upper Bound)      & 0.815& 0.631  &  0.649 &  0.675  & 0.707   \\ \cline{2-7}
                                    & Regression-EM  & 0.815& 0.629  & 0.648  & 0.674   & 0.705   \\ \cline{2-7}
                                    & Randomization~\cite{DBLP:conf/wsdm/JoachimsSS17}       
                                                     &0.814& 0.628  & 0.644  & 0.672  & 0.707   \\ \cline{2-7}  
                                    & Click Data (Lower Bound)    &0.811& 0.614  & 0.629  & 0.658  & 0.697   \\ \hline

\end{tabular}
\end{table*}

\subsubsection{Baseline Methods} 

We made comprehensive comparisons between our method and the baselines. The baselines were created by combining the state-of-the-art debiasing methods and learning-to-rank algorithms. There were six debiasing methods. To make fair comparison, we used click model to generate 165660 query sessions as training dataset, and utilized the same dataset for all debiasing methods. All the hyper-parameters of the baseline models were the same as those in the original papers.

\textbf{Randomization}: The method, proposed by Joachims et al.~\cite{DBLP:conf/wsdm/JoachimsSS17}, uses randomization to infer the observation probabilities as position biases. We randomly shuffled the rank lists and then estimated the position biases as in~\cite{DBLP:conf/sigir/AiBLGC18}.

\textbf{Regression-EM}: The method, proposed by Wang et al.~\cite{DBLP:conf/wsdm/WangGBMN18}, directly estimates the position biases using a regression-EM model implemented by GBDT. 

\textbf{Dual Learning Algorithm}: The method, proposed by Ai et al.~\cite{DBLP:conf/sigir/AiBLGC18}, jointly learns a ranker and conducts debiasing of click data. The algorithm implements both the ranking model and the debiasing model as deep neural networks.

\textbf{Pairwise Debiasing}: Our proposed debiasing method, which is combined with LambdaMART. 
In this experiment, we set the hyper-parameter $p$ as 0 by default. As explained below, a further experiment was conducted with different values of $p$.

\textbf{Click Data}: In this method, the raw click data without debiasing is used to train a ranker, whose performance is considered as a lower bound.

\textbf{Labeled Data}: In this method, human annotated relevance labels without any bias are used as data for training of ranker, whose performance is considered as an upper bound.

There were three learning-to-rank algorithms.

\textbf{DNN}: A deep neural network was implemented as a ranker, as in ~\cite{DBLP:conf/sigir/AiBLGC18}. We directly used the code provided by Ai et al.\footnote{~\url{https://github.com/QingyaoAi/Dual-Learning-Algorithm-for-Unbiased-Learning-to-Rank}}.

\textbf{RankSVM}: We directly used the Unbiased RankSVM Software provided by Joachims et al.\footnote{~\url{https://www.cs.cornell.edu/people/tj/svm_light/svm_proprank.html}}, with hyper-parameter $C$ being 200. 

\textbf{LambdaMART}: We implemented Unbiased LambdaMART by modifying the LambdaMART tool in LightGBM~\cite{DBLP:conf/nips/KeMFWCMYL17}. We utilized the default hyper-parameters of the tool. The total number of trees was 300, learning rate was 0.05, number of leaves for one tree was 31, feature fraction was 0.9, and bagging fraction was 0.9.

In summary, there were 13 baselines to compare with our proposed Unbiased LambdaMART algorithm. Note that Dual Learning Algorithm and DNN are tightly coupled. We did not combine Pairwise Debiasing with RankSVM and DNN, as it is beyond the scope of this paper.

\subsubsection{Experimental Results} 

Table ~\ref{tab:simulate} summarizes the results. We can see that our method of Unbiased LambdaMART (LambdaMART + Pairwise Debiasing) significantly outperforms all the other baseline methods. The results of Regression-EM and Dual Learning Algorithm are compariable with those reported in the original papers. In particular, we have the following findings.
\begin{itemize}
       \item Our method of LambdaMART+Pairwise Debiasing (Unbiased LambdaMART) achieves better performances than all the state-of-the-art methods in terms of all measures. For example, in terms of NDCG@1, our method outperforms LambdaMART+Regression-EM by 3.2\%, outperforms DNN+Dual Learning by 4.3\%, and outperforms RankSVM+Randomization by 8.9\%. 
       
       \item Pairwise Debiasing works better than the other debiasing methods. When combined with LambdaMART, Pairwise Debiasing outperforms Regression-EM by 3.2\%, outperforms Randomization by 4.8\% in terms of NDCG@1.
  
       \item LambdaMART trained with human labeled data achieves the best performance (upper bound). An unbiased learning-to-rank algorithm can still not beat it. This indicates that there is still room for improvement in unbiased learning-to-rank.  
      
       \item When trained with Click Data, the performance of LambdaMART decreases significantly and gets closer to those of RankSVM and DNN. This implies that a sophisticated algorithm like LambdaMART is more sensitive to position bias. 
 
\end{itemize}

\begin{figure}[t]
\centering   
\includegraphics [width=0.45\textwidth]{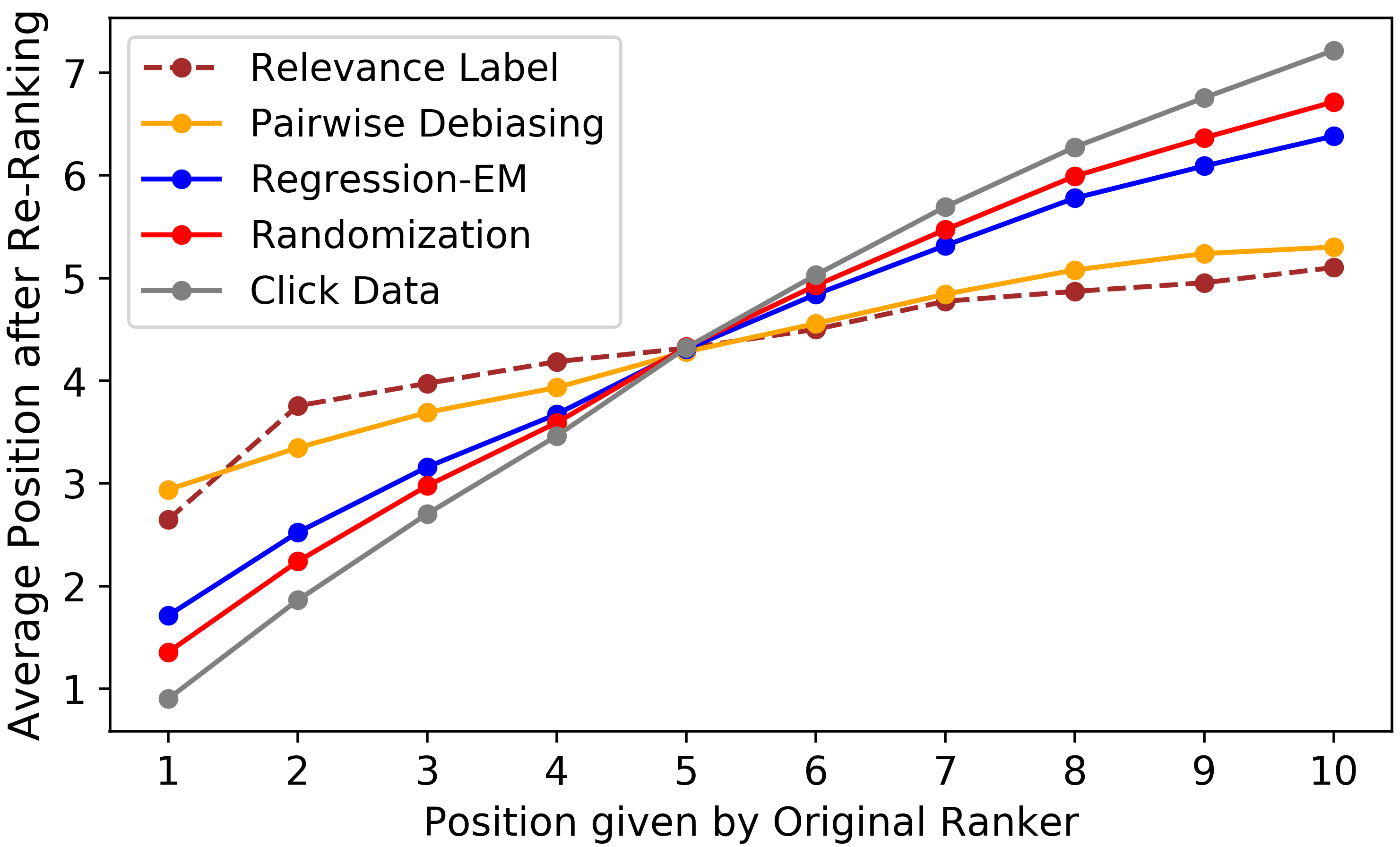} 
\caption{Average positions after re-ranking of documents at each original position by different debiasing methods with LamdbaMART.} 
 \label{fig:diff} 
\end{figure}

\begin{figure}[t]
\centering   
\includegraphics [width=0.47\textwidth]{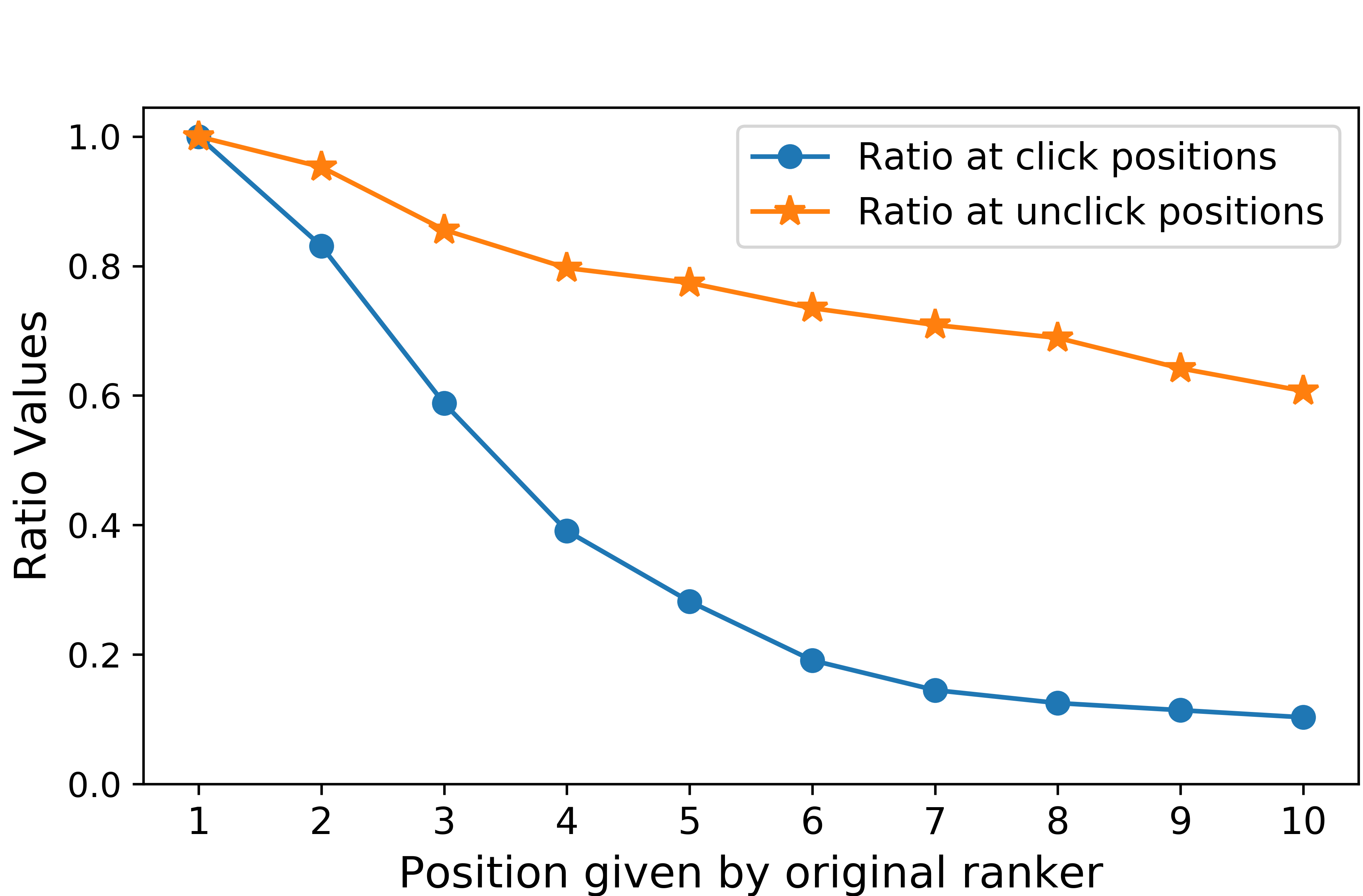} 
\caption{Position biases (ratios) at click and unclick positions estimated by Unbiased LambdaMART.} 
 \label{fig:curve} 
\end{figure}

\begin{figure*}[t!]
    \centering
    \subfigure[Performance on click data generated by Cascade Model]
    {
        \begin{minipage}{0.45\linewidth}
            \includegraphics[width=7.5cm]{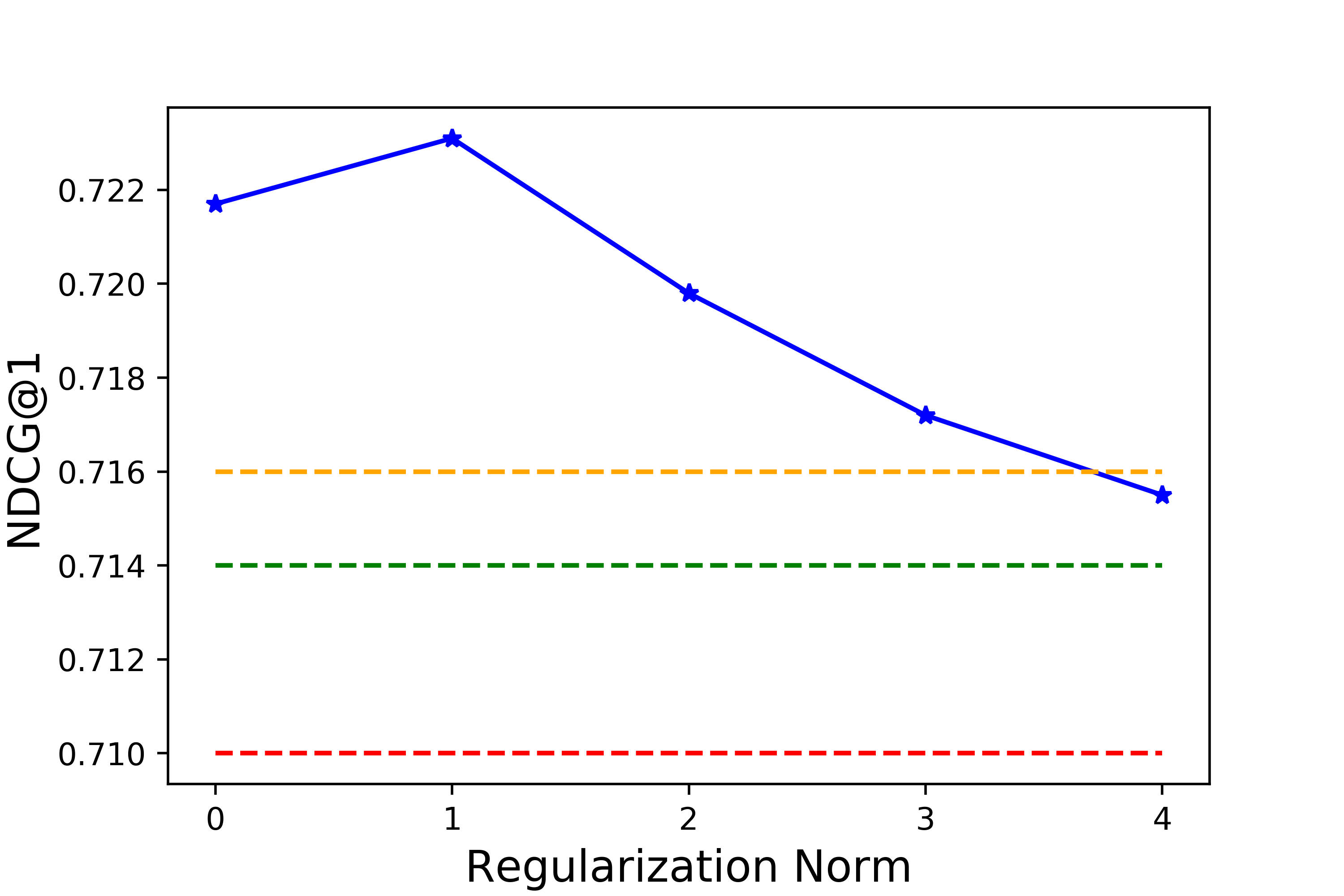}
            \includegraphics[width=7.5cm]{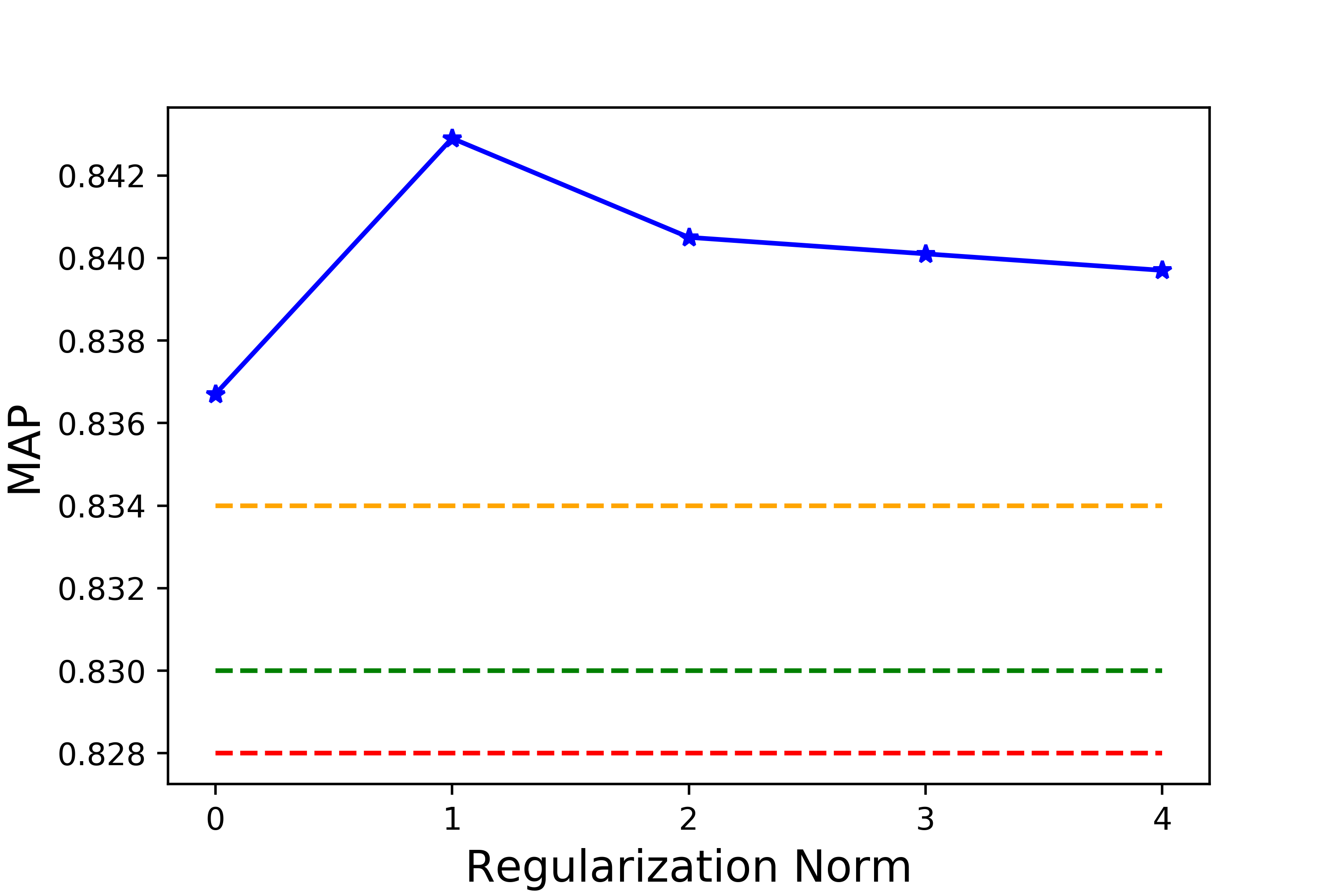}
        \end{minipage}
    }
    ~ 
    \subfigure[Performance on click data generated by Position Based Model]
    {
        \begin{minipage}{0.45\linewidth}
            \includegraphics[width=7.5cm]{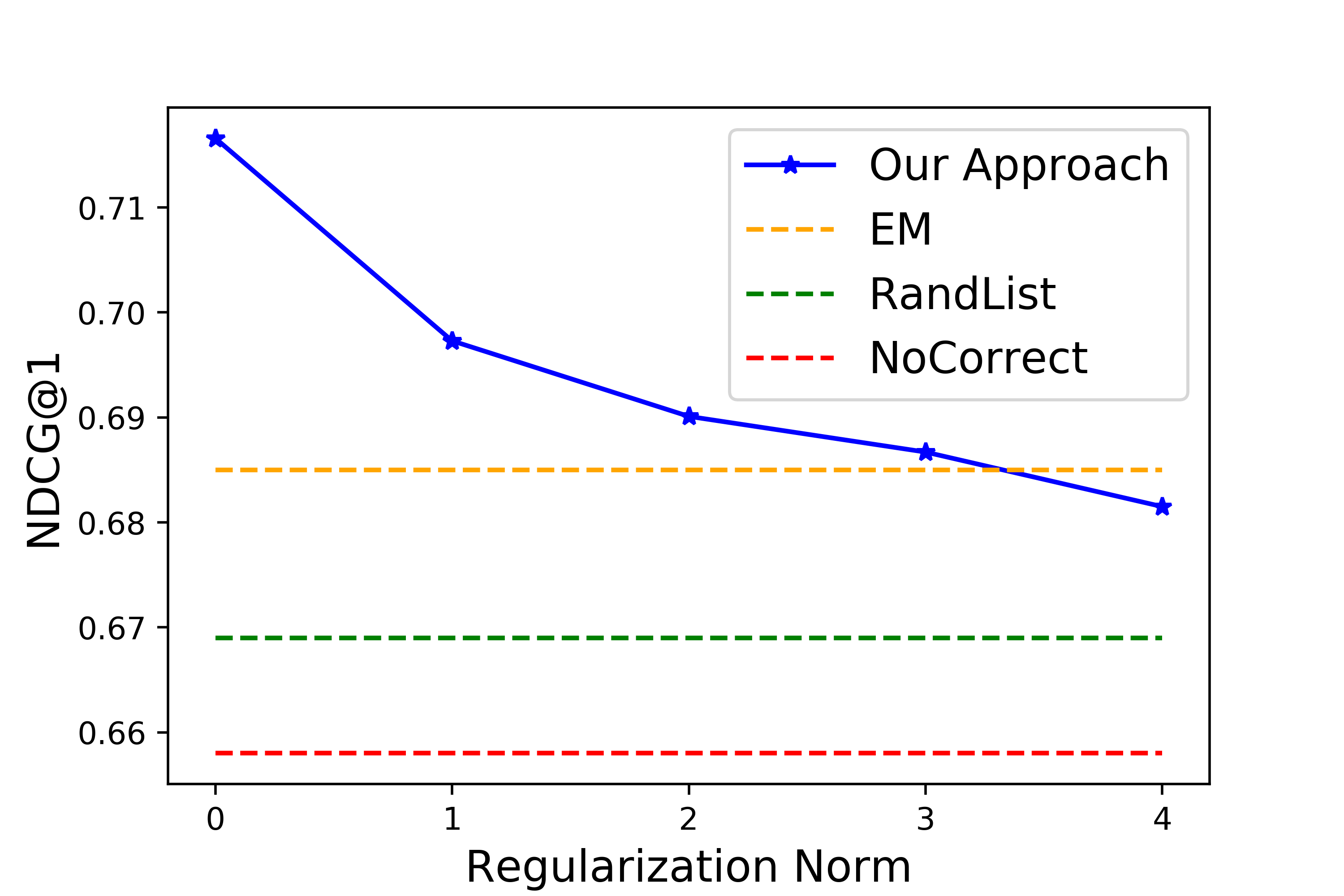}
            \includegraphics[width=7.5cm]{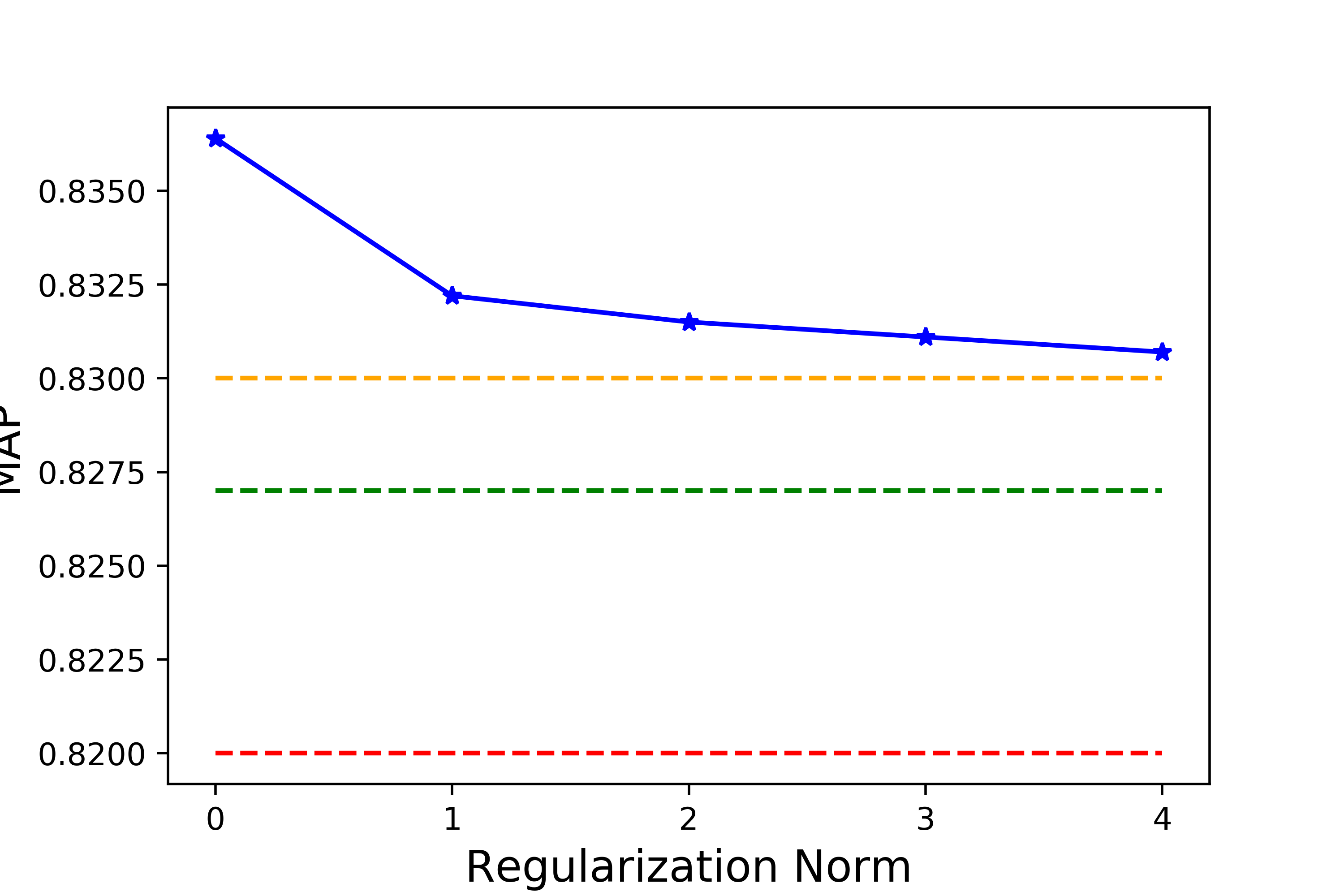}
        \end{minipage}
    }
    \caption{Performances of LambdaMART versus regularization norms by different debiasing methods, when click data is generated by two different click models. }
    \label{fig:exp}
\end{figure*}

\subsection{Empirical Analysis}

We conducted additional experiments to investigate the effectiveness, generalizability, and robustness of Unbiased LambdaMART.


\subsubsection{Effectiveness of Unbiased LambdaMART}

We investigated whether the performance improvement by Unbiased LambdaMART is indeed from reduction of position bias. 

We first identified the documents at each position given by the original ranker.
We then calculated the average positions of the documents at each original position after re-ranking by Pairwise Debiasing and the other debiasing methods, combined with LambdaMART. We also calculated the average positions of the documents after re-ranking by their relevance labels, which is the ground truth. Ideally, the average positions by the debiasing methods should get close to the average positions by the relevance labels. Figure \ref{fig:diff} shows the results.

One can see that the curve of LambdaMART + Click Data (in grey) is away from that of relevance labels or ground truth (in brown), indicating that directly using click data without debiasing can be problematic. The curve of Pairwise Debiasing (in orange) is the closest to the curve of relevance labels, indicating that the performance enhancement by Pairwise Debiasing is indeed from effective debiasing.

Figure \ref{fig:curve} shows the normalized (relative) position biases for click and unclick positions given by Unbiased LambdaMART. The result indicates that both the position biases at click positions and position biases at unclick positions monotonically decrease, while the former decrease at a faster rate than the latter. The result exhibits how Unbiased LambdaMART can reduce position biases in the pairwise setting.

\begin{figure*}[t!]
    \centering
    \subfigure
    {
    \begin{minipage}{0.47\linewidth}
    \includegraphics[width=7.8cm]{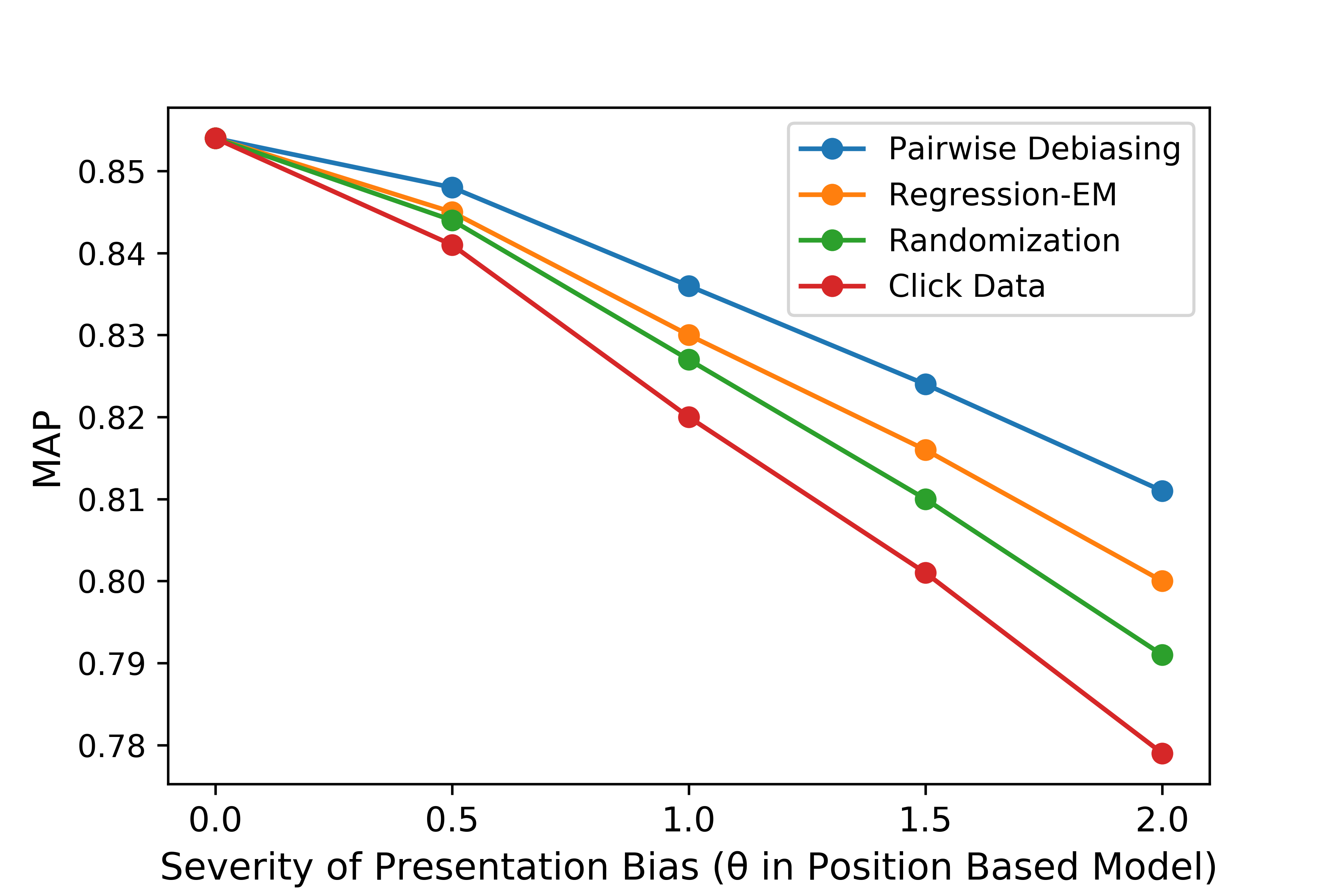}
    \end{minipage}
    }
    ~ 
    \subfigure
    {
    \begin{minipage}{0.47\linewidth}
    \includegraphics[width=7.8cm]{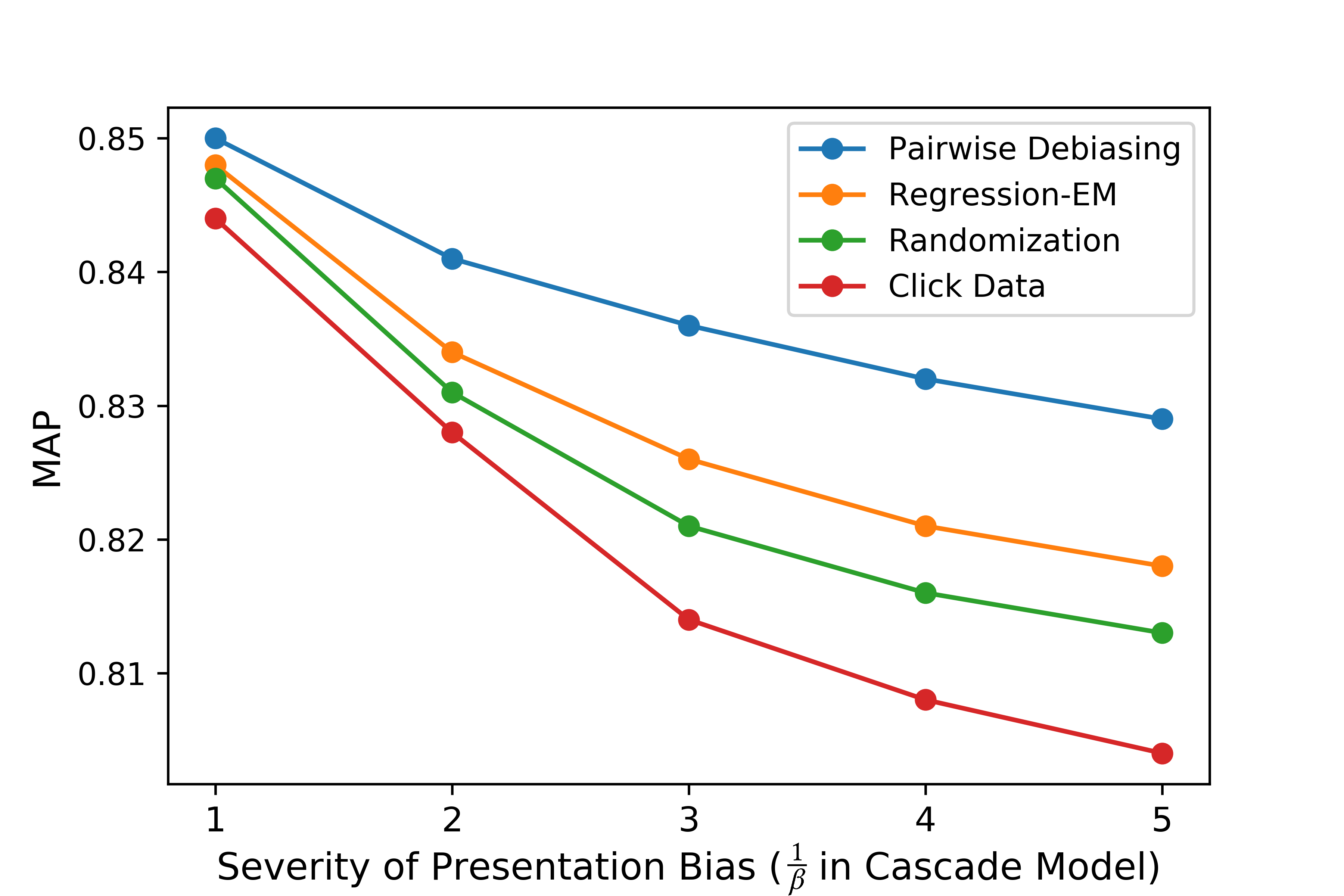}
    \end{minipage}
    }
    \caption{Performances of Pairwise Debiasing against other debiasing methods with different degrees of position bias.}
    \label{fig:robust}
\end{figure*}

\subsubsection{Generalizability of Unbiased LambdaMART}

The Position Based Model (PBM) assumes that the bias of a document only depends on its position, which is an approximation of user click behavior in practice. 
The Cascade Model ~\cite{DBLP:conf/sigir/DupretP08}, on the other hand, assumes that the user browses the search results in a sequential order from top to bottom, which may more precisely model user behavior. We therefore analyzed the generalizability of Unbiased LambdaMART by using simulated click data from both Position Based Model and Cascade Model, and studied whether regularization of position bias, i.e., hyper-parameter $p$, affects performance.

We used a variant of Cascade Model which is similar to Dynamic Bayesian Model in \cite{DBLP:conf/www/ChapelleZ09}. There is a probability $\phi$ that the user is satisfied with the result after clicking the document. If the user is satisfied, he~\slash~she will stop searching; and otherwise, there is a probability $\beta$ that he~\slash~she will examine the next result and there is a probability $1-\beta$ that he~\slash~she will stop searching. Obviously, the smaller $\beta$ indicates that the user will have a smaller probability to continue reading, which means a more severe position bias. In our experiment, we set $\phi$ as half of the relevance probability and used the default value of $\beta$ i.e., 0.5.

We compared Unbiased LambdaMART (LambdaMART + Pairwise Debiasing) with LambdaMART + two different debiasing methods, Regression-EM and Randomization, and also Click Data without debiasing on the two datasets. Again, we found that Unbiased LambdaMART significantly outperforms the baselines, indicating that Pairwise Debiasing is indeed an effective method.

Figure \ref{fig:exp} shows the results of the methods in terms of NDCG@1 and MAP, where we choose NDCG@1 as representative of NDCG scores. For Unbiased LambdaMART, it shows the results under different hyper-parameter values. We can see that Unbiased LambdaMART is superior to all the three baselines on both datasets generated by Position Based Model and Cascade Model. We can also see that in general to achieve high performance the value of $p$ in $L_p$ regularization should not be so high. For the dataset generated by Cascade Model, the performance with $L_1$ regularization is better than that with $L_0$ regularization. It indicates that when the data violates its assumption, Unbiased LambdaMART can still learn a reliable model with a higher order of regularization. 

\begin{figure}[h!]
\centering   
\includegraphics [width=0.47\textwidth]{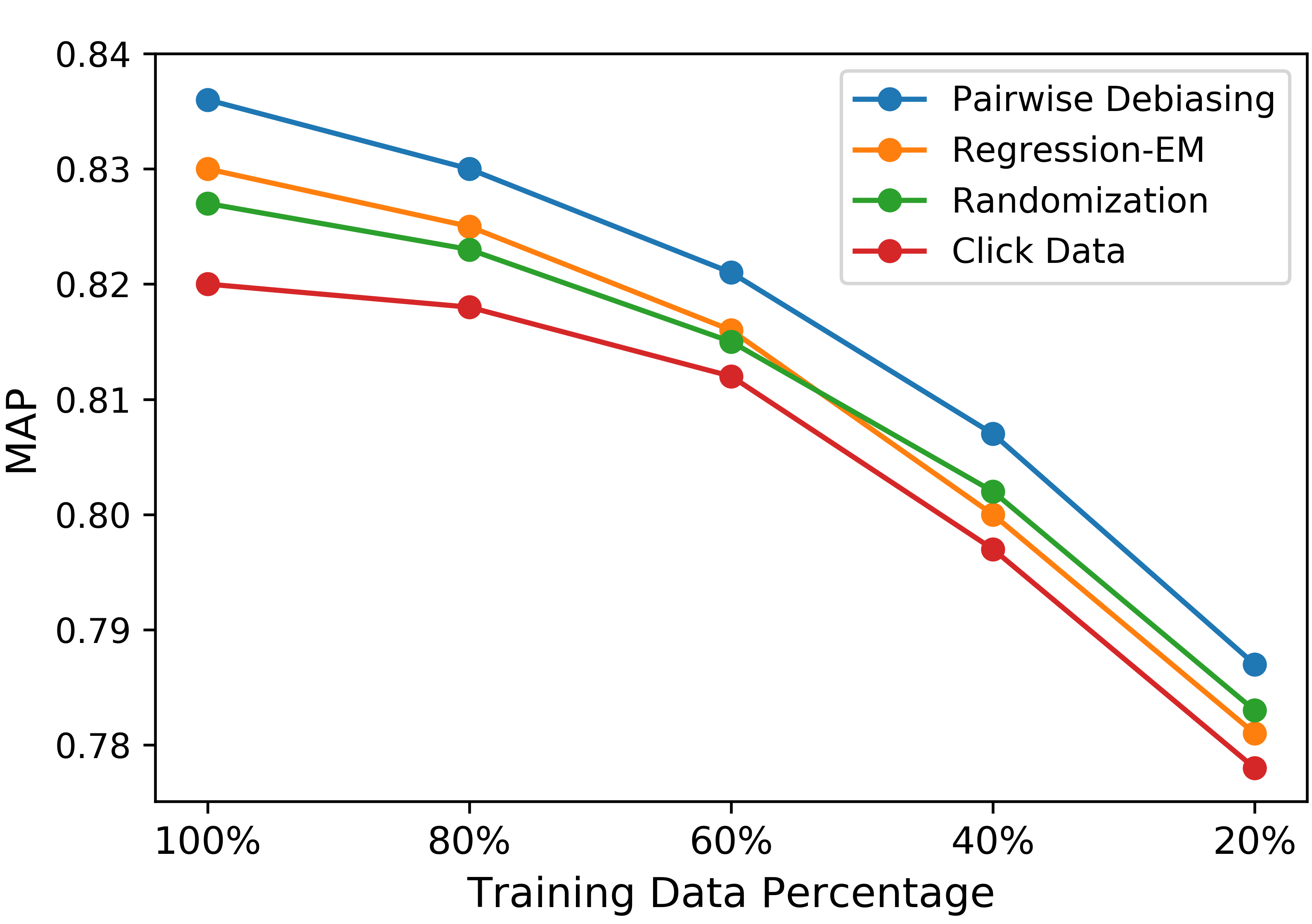} 
\caption{Performances of Pairwise Debiasing against other debiasing methods with different sizes of training data.} 

 \label{fig:perc} 
\end{figure}

\subsubsection{Robustness of Unbiased LambdaMART}

We further evaluated the robustness of Unbiased LambdaMART under different degrees of position bias.

In the above experiments, we only tested the performance of Unbiased LambdaMART with click data generated from a single click model, i.e., $\theta$ as $1$ for Position Based Model and $\beta$ as $0.5$ for Cascade Model. Therefore, here we set the two hyper-parameters to different values and examined whether Unbiased LambdaMART can still work equally well. 

Figure \ref{fig:robust} shows the results in terms of NDCG@1 with different degrees of position bias. The results in terms of other measures have similar trends. When $\theta$ in Position Based Model equals $0$, and $\beta$ in Cascade Model equals $1$, there is no position bias. The results of all debiasing methods are similar to that of using click data only. As we add more position bias, i.e., $\theta$ increases and $\beta$ decreases, the performances of all the debiasing methods decrease dramatically. However, under all settings Unbiased LambdaMART can get less affected by position bias and consistently maintain the best results. This indicates that Unbiased LambdaMART is robust to different degrees of position bias. 

Next, we investigate the robustness of Unbiased LambdaMart under different sizes of training data. We randomly selected a subset of training data, (i.e., 20\% - 100\%) to generate different sizes of click datasets, and used the datasets to evaluate the performances of LambdaMART with different debiasing methods. To make fair comparison, we used the same subsets of training data for running of the Randomization and Regression-EM algorithm.

As shown in Figure \ref{fig:perc}, when the size of training data decreases, the improvements obtained by the debiasing methods also decrease. The reason seems to be that the position bias estimated from insufficient training data is not accurate, which can hurt the performances of the debiasing methods. In contrast, Unbiased LambdaMART, which adopts a joint training mechanism, can still achieve the best performances in such cases. When the data size increases from 80\% to 100\%, the performance enhancement of LambdaMART + Click Data is quite small, while the performance enhancements of the debiasing methods are much larger. This result is in accordance with the observation reported in\cite{DBLP:conf/wsdm/JoachimsSS17}, that simply increasing the amount of biased training data cannot help build a reliable ranking model, but after debiasing it is possible to learn a better ranker with more training data. The experiment shows that Unbiased LambdaMART can still work well even with limited training data, and it can consistently increase its performances as training data increases. 

\subsection{A/B Testing at Commercial Search Engine}
We further evaluated the performance of Unbiased LambdaMART by deploying it at the search engine of Jinri Toutiao, a commercial news recommendation app in China with over 100 million daily active users. We trained two rankers with Unbiased LambdaMART and LambdaMART + Click Data using click data of approximately 19.6 million query sessions collected over two days at the search engine.
Then we deployed the two rankers at the search system to conduct A/B testing. The A/B testing was carried out for 16 days. In each experiment group, the ranker was randomly assigned approximately 1.5 million queries per day.

In the online environment, we observed that different users have quite different click behaviors. It appeared to be necessary to have a tighter control on debiasing. We therefore set the hyper-parameter $p$ as 1, i.e., we conducted $L_1$ regularization to impose a stronger regularization on the position biases. We validated the correctness of this hyper-parameter selection on a small set of relevance dataset.

We compared the results of the two rankers in terms of first click ratios, which are the percentages of sessions having first clicks at top 1,3,5 positions among all sessions. A ranker with higher first click ratios should have better performance. 

\begin{table}[t]
\Large
\caption{Relative increases of first click ratios by Unbiased LambdaMART in online A/B testing.} 
\label{tab:ab} 
\begin{tabular}{c|c|c|c}
 \hline
  Measure & Click@1 & Click@3 & Click@5 \\ \hline \hline
Increase & 2.64\%       & 1.21\%       & 0.80\%     \\ \hline
P-value       & 0.001        & 0.004        & 0.023      \\ \hline
\end{tabular}
\end{table}

\begin{table}[t]
\Large
\caption{Human assessors' evaluation on results of same queries ranked at top five positions by the two rankers.}
\label{tab:annot} 
\begin{tabular}{c||c|c|c}
\hline
\multirow{2}{*}{\begin{tabular}[c]{@{}l@{}}Unbiased LambdaMART \\ \emph{vs.} LambdaMart + Click\end{tabular}} & Win & Same & Loss \\ \cline{2-4} 
                                                                                                      & 21     & 68  & 11     \\ \hline
\end{tabular}
\end{table}

As shown in Table ~\ref{tab:ab}, Unbiased LambdaMART can significantly outperform LambdaMART + Click Data in terms of first click ratios at the A/B Testing. It increases the first click ratios at positions 1,3,5 by 2.64\%, 1.21\% and 0.80\%, respectively, which are all statistically significant (p-values < 0.05). It indicates that Unbiased LambdaMART can make significantly better relevance ranking with its debiasing capability.

We next asked human assessors to evaluate the results of the two rankers.
We collected all the different results of the same queries given by the two rankers during the A/B testing period, presented the results to the assessors randomly side-by-side,
and asked assessors to judge which results are better. They categorized the results at the top five positions of 100 randomly chosen queries into three categories, i.e., `Win', `Same' and `Loss'. 

As shown in Table ~\ref{tab:annot}, the win/loss ratio of Unbiased LambdaMart over LambdaMart + Click Data is as high as 1.91, indicating that Unbiased LambdaMART is indeed effective as an unbiased learning-to-rank algorithm. 

\section{Conclusion}\label{sec:conclusion}
In this paper, we have proposed a general framework for pairwise unbiased learning-to-rank, including the extended inverse propensity weighting (IPW) principle. We have also proposed a method called Pairwise Debiasing to jointly estimate position biases and train a ranker by directly optimizing a same objective function within the framework. We develop a new algorithm called Unbiased LambdaMART as application of the method. Experimental results show that Unbiased LambdaMART achieves significantly better results than the existing methods on a benchmark dataset, and is effective in relevance ranking at a real-world search system.

There are several items to work on in the future. We plan to apply Pairwise Debiasing to other pairwise learning-to-rank algorithms. We also consider developing a more general debiasing method that can deal with not only position bias but also other types of bias such as presentation bias. More theoretical analysis on unbiased pairwise learning-to-rank is also necessary.

\bibliographystyle{acm}\balance
\bibliography{www}

\begin{thebibliography}{10}

\bibitem{DBLP:conf/sigir/AiBLGC18}
{\sc Ai, Q., Bi, K., Luo, C., Guo, J., and Croft, W.~B.}
\newblock Unbiased learning to rank with unbiased propensity estimation.
\newblock In {\em The 41st International {ACM} {SIGIR} Conference on Research
  {\&} Development in Information Retrieval, {SIGIR} 2018, Ann Arbor, MI, USA,
  July 08-12, 2018\/} (2018), pp.~385--394.

\bibitem{DBLP:conf/cikm/AiMLC18}
{\sc Ai, Q., Mao, J., Liu, Y., and Croft, W.~B.}
\newblock Unbiased learning to rank: Theory and practice.
\newblock In {\em Proceedings of the 27th ACM International Conference on
  Information and Knowledge Management, {CIKM} 2018\/} (2018), {ACM},
  pp.~2305--2306.

\bibitem{DBLP:conf/www/BorisovMRS16}
{\sc Borisov, A., Markov, I., de~Rijke, M., and Serdyukov, P.}
\newblock A neural click model for web search.
\newblock In {\em Proceedings of the 25th International Conference on World
  Wide Web, {WWW} 2016, Montreal, Canada, April 11 - 15, 2016\/} (2016),
  pp.~531--541.

\bibitem{from-ranknet-to-lambdarank-to-lambdamart-an-overview}
{\sc Burges, C.~J.}
\newblock From ranknet to lambdarank to lambdamart: An overview.
\newblock Tech. rep., June 2010.

\bibitem{DBLP:conf/nips/BurgesRL06}
{\sc Burges, C. J.~C., Ragno, R., and Le, Q.~V.}
\newblock Learning to rank with nonsmooth cost functions.
\newblock In {\em Proceedings of the 20th Annual Conference on Neural
  Information Processing Systems, {NIPS} 2006, Vancouver, British Columbia,
  Canada, December 4-7, 2006\/} (2006), pp.~193--200.

\bibitem{Cao:2006:ARS:1148170.1148205}
{\sc Cao, Y., Xu, J., Liu, T.-Y., Li, H., Huang, Y., and Hon, H.-W.}
\newblock Adapting ranking svm to document retrieval.
\newblock In {\em Proceedings of the 29th Annual International ACM SIGIR
  Conference on Research and Development in Information Retrieval, {SIGIR}
  2006\/} (New York, NY, USA, 2006), ACM, pp.~186--193.

\bibitem{DBLP:journals/jmlr/ChapelleC11}
{\sc Chapelle, O., and Chang, Y.}
\newblock Yahoo! learning to rank challenge overview.
\newblock In {\em Proceedings of the Yahoo! Learning to Rank Challenge, held at
  {ICML} 2010, Haifa, Israel, June 25, 2010\/} (2011), pp.~1--24.

\bibitem{DBLP:conf/www/ChapelleZ09}
{\sc Chapelle, O., and Zhang, Y.}
\newblock A dynamic bayesian network click model for web search ranking.
\newblock In {\em Proceedings of the 18th International Conference on World
  Wide Web, {WWW} 2009, Madrid, Spain, April 20-24, 2009\/} (2009), pp.~1--10.

\bibitem{DBLP:conf/wsdm/CraswellZTR08}
{\sc Craswell, N., Zoeter, O., Taylor, M.~J., and Ramsey, B.}
\newblock An experimental comparison of click position-bias models.
\newblock In {\em Proceedings of the International Conference on Web Search and
  Web Data Mining, {WSDM} 2008, Palo Alto, California, USA, February 11-12,
  2008\/} (2008), pp.~87--94.

\bibitem{DBLP:conf/sigir/DupretP08}
{\sc Dupret, G., and Piwowarski, B.}
\newblock A user browsing model to predict search engine click data from past
  observations.
\newblock In {\em Proceedings of the 31st Annual International {ACM} {SIGIR}
  Conference on Research and Development in Information Retrieval, {SIGIR}
  2008, Singapore, July 20-24, 2008\/} (2008), pp.~331--338.

\bibitem{Friedman00greedyfunction}
{\sc Friedman, J.~H.}
\newblock Greedy function approximation: A gradient boosting machine.
\newblock {\em Annals of Statistics 29\/} (2000), 1189--1232.

\bibitem{DBLP:journals/tois/JarvelinK02}
{\sc J{\"{a}}rvelin, K., and Kek{\"{a}}l{\"{a}}inen, J.}
\newblock Cumulated gain-based evaluation of {IR} techniques.
\newblock {\em {ACM} Trans. Inf. Syst. 20}, 4 (2002), 422--446.

\bibitem{DBLP:conf/kdd/Joachims02}
{\sc Joachims, T.}
\newblock Optimizing search engines using clickthrough data.
\newblock In {\em Proceedings of the Eighth {ACM} {SIGKDD} International
  Conference on Knowledge Discovery and Data Mining, July 23-26, 2002,
  Edmonton, Alberta, Canada\/} (2002), pp.~133--142.

\bibitem{DBLP:conf/sigir/JoachimsGPHG05}
{\sc Joachims, T., Granka, L.~A., Pan, B., Hembrooke, H., and Gay, G.}
\newblock Accurately interpreting clickthrough data as implicit feedback.
\newblock In {\em {SIGIR} 2005: Proceedings of the 28th Annual International
  {ACM} {SIGIR} Conference on Research and Development in Information
  Retrieval, Salvador, Brazil, August 15-19, 2005\/} (2005), pp.~154--161.

\bibitem{DBLP:conf/wsdm/JoachimsSS17}
{\sc Joachims, T., Swaminathan, A., and Schnabel, T.}
\newblock Unbiased learning-to-rank with biased feedback.
\newblock In {\em Proceedings of the Tenth {ACM} International Conference on
  Web Search and Data Mining, {WSDM} 2017, Cambridge, United Kingdom, February
  6-10, 2017\/} (2017), pp.~781--789.

\bibitem{DBLP:conf/nips/KeMFWCMYL17}
{\sc Ke, G., Meng, Q., Finley, T., Wang, T., Chen, W., Ma, W., Ye, Q., and Liu,
  T.}
\newblock Lightgbm: {A} highly efficient gradient boosting decision tree.
\newblock In {\em Advances in Neural Information Processing Systems 30: Annual
  Conference on Neural Information Processing Systems, {NIPS} 2017, 4-9
  December 2017, Long Beach, CA, {USA}\/} (2017), pp.~3149--3157.

\bibitem{DBLP:conf/icml/KvetonSWA15}
{\sc Kveton, B., Szepesvari, C., Wen, Z., and Ashkan, A.}
\newblock Cascading bandits: Learning to rank in the cascade model.
\newblock In {\em Proceedings of the 32nd International Conference on Machine
  Learning, {ICML} 2015, Lille, France, 6-11 July 2015\/} (2015), pp.~767--776.

\bibitem{DBLP:journals/ieicet/Li11}
{\sc Li, H.}
\newblock A short introduction to learning to rank.
\newblock {\em {IEICE} Transactions 94-D}, 10 (2011), 1854--1862.

\bibitem{DBLP:series/synthesis/2014Li}
{\sc Li, H.}
\newblock {\em Learning to Rank for Information Retrieval and Natural Language
  Processing, Second Edition}.
\newblock Synthesis Lectures on Human Language Technologies. Morgan {\&}
  Claypool Publishers, 2014.

\bibitem{DBLP:conf/kdd/LiAKMVW18}
{\sc Li, S., Abbasi-Yadkori, Y., Kveton, B., Muthukrishnan, S., Vinay, V., and
  Wen, Z.}
\newblock Offline evaluation of ranking policies with click models.
\newblock In {\em Proceedings of the 24th ACM SIGKDD International Conference
  on Knowledge Discovery and Data Mining, {KDD}\/} (2018), {ACM},
  pp.~1685--1694.

\bibitem{DBLP:journals/ftir/Liu09}
{\sc Liu, T.}
\newblock Learning to rank for information retrieval.
\newblock {\em Foundations and Trends in Information Retrieval 3}, 3 (2009),
  225--331.

\bibitem{DBLP:conf/www/RichardsonDR07}
{\sc Richardson, M., Dominowska, E., and Ragno, R.}
\newblock Predicting clicks: estimating the click-through rate for new ads.
\newblock In {\em Proceedings of the 16th International Conference on World
  Wide Web, {WWW} 2007, Banff, Alberta, Canada, May 8-12, 2007\/} (2007),
  pp.~521--530.

\bibitem{rose:rubi:cent:1983}
{\sc Rosenbaum, P.~R., and Rubin, D.~B.}
\newblock The central role of the propensity score in observational studies for
  causal effects.
\newblock {\em Biometrika 70\/} (1983), 41--55.

\bibitem{DBLP:conf/sigir/WangBMN16}
{\sc Wang, X., Bendersky, M., Metzler, D., and Najork, M.}
\newblock Learning to rank with selection bias in personal search.
\newblock In {\em Proceedings of the 39th International {ACM} {SIGIR}
  conference on Research and Development in Information Retrieval, {SIGIR}
  2016, Pisa, Italy, July 17-21, 2016\/} (2016), pp.~115--124.

\bibitem{DBLP:conf/wsdm/WangGBMN18}
{\sc Wang, X., Golbandi, N., Bendersky, M., Metzler, D., and Najork, M.}
\newblock Position bias estimation for unbiased learning to rank in personal
  search.
\newblock In {\em Proceedings of the Eleventh {ACM} International Conference on
  Web Search and Data Mining, {WSDM} 2018, Marina Del Rey, CA, USA, February
  5-9, 2018\/} (2018), pp.~610--618.

\bibitem{DBLP:journals/ir/WuBSG10}
{\sc Wu, Q., Burges, C. J.~C., Svore, K.~M., and Gao, J.}
\newblock Adapting boosting for information retrieval measures.
\newblock {\em Inf. Retr. 13}, 3 (2010), 254--270.

\bibitem{DBLP:conf/www/YuePR10}
{\sc Yue, Y., Patel, R., and Roehrig, H.}
\newblock Beyond position bias: examining result attractiveness as a source of
  presentation bias in clickthrough data.
\newblock In {\em Proceedings of the 19th International Conference on World
  Wide Web, {WWW} 2010, Raleigh, North Carolina, USA, April 26-30, 2010\/}
  (2010), pp.~1011--1018.

\end{thebibliography}

\end{document}